\documentclass[graybox]{svmult}

\usepackage{mathptmx}       
\usepackage{helvet}         
\usepackage{courier}        
\usepackage{type1cm}        
\usepackage{makeidx}         
\usepackage{graphicx}        
\usepackage{multicol}        
\usepackage[bottom]{footmisc}

\makeindex             

\def\teff{\ifmmode T_{\rm eff} \else $T_{\mathrm{eff}}$\fi}

\def\ltsima{$\buildrel<\over\sim$}
\def\lsim{\lower.5ex\hbox{\ltsima}}

\newcommand{\hii}{H~{\sc ii}}
\newcommand{\ha}{\ifmmode {\rm H}\alpha \else H$\alpha$\fi}
\newcommand{\hb}{\ifmmode {\rm H}\beta \else H$\beta$\fi}
\newcommand{\lya}{\ifmmode {\rm Ly}\alpha \else Ly$\alpha$\fi}

\newcommand{\ebv}{\ifmmode E_{\rm B-V} \else $E_{\rm B-V}$\fi}
\newcommand{\av}{\ifmmode A_{\rm V} \else $A_{\rm V}$\fi}

\def\msun{\ifmmode M_{\odot} \else M$_{\odot}$\fi}
\def\msunyr{\ifmmode M_{\odot} {\rm yr}^{-1} \else M$_{\odot}$ yr$^{-1}$\fi}
\def\zsun{\ifmmode Z_{\odot} \else Z$_{\odot}$\fi}

\def\lsun{\ifmmode L_{\odot} \else L$_{\odot}$\fi}

\def\mup{\ifmmode M_{\rm up} \else M$_{\rm up}$\fi}
\def\mlow{\ifmmode M_{\rm low} \else M$_{\rm low}$\fi}

\newcommand{\oh}{\ifmmode 12 + \log({\rm O/H}) \else$12 + \log({\rm
O/H})$\fi}
\def\flyf{\ifmmode f_{\rm Lyf} \else $f_{\rm Lyf}$\fi}
\def\pz{\ifmmode P(z) \else $P(z)$\fi}
\def\ki2{\ifmmode \chi^2 \else $\chi^2$\fi}
\def\zphot{\ifmmode z_{\rm phot} \else $z_{\rm phot}$\fi}

\newcommand{\xphot}{\ifmmode x_\gamma \else $v_\gamma$\fi}
\newcommand{\xobs}{\ifmmode x_{\rm obs} \else $x_{\rm obs}$\fi}
\newcommand{\xcmf}{\ifmmode x_{\rm CMF} \else $x_{\rm CMF}$\fi}
\newcommand{\vexp}{\ifmmode V_{\rm exp} \else $V_{\rm exp}$\fi}
\newcommand{\vmax}{\ifmmode V_{\rm max} \else $V_{\rm max}$\fi}
\newcommand{\nh}{\ifmmode N_{\rm HI} \else $N_{\rm HI}$\fi}
\newcommand{\dv}{\ifmmode \Delta v({\rm em-abs}) \else $\Delta v({\rm em}-{\rm abs})$\fi}

\def\fesc{\ifmmode f_{\rm esc} \else $f_{\rm esc}$\fi}

\def\frellya{\ifmmode f^{\rm rel}_{\rm{Ly}\alpha} \else $f^{\rm rel}_{\rm{Ly}\alpha}$\fi}
\def\wlya{$W_{\rm{Ly}\alpha}$}

\newcommand{\mstar}{\ifmmode M_\star \else $M_\star$\fi}
\newcommand{\mdust}{\ifmmode M_d \else $M_d$\fi}
\newcommand{\muv}{\ifmmode M_{1500} \else $M_{1500}$\fi}
\newcommand{\luv}{\ifmmode L_{\rm UV} \else $L_{\rm UV}$\fi}
\newcommand{\lir}{\ifmmode L_{\rm IR} \else $L_{\rm IR}$\fi}
\newcommand{\lbol}{\ifmmode L_{\rm bol} \else $L_{\rm bol}$\fi}
\newcommand{\liruv}{\ifmmode L_{\rm IR+UV} \else $L_{\rm IR+UV}$\fi}
\newcommand{\liroveruv}{\ifmmode L_{\rm IR}/L_{\rm UV} \else $L_{\rm IR}/L_{\rm UV}$\fi}
\newcommand{\nlyc}{\ifmmode N_{\rm Lyc} \else $N_{\rm Lyc} $\fi}
\newcommand{\rholyc}{\ifmmode \rho_{\rm Lyc} \else $\rho_{\rm Lyc} $\fi}
\newcommand{\auv}{\ifmmode  A_{\rm UV} \else $A_{\rm UV}$\fi}

\newcommand{\Cii}{\ifmmode [{\rm CII}] \else $[{\rm CII}]$\fi}
\newcommand{\lcii}{\ifmmode L_{[\rm CII]} \else $L_{[\rm CII]}$\fi}
\newcommand{\lirngc}{\ifmmode L_{\rm IR}^{\rm N6946} \else $L_{\rm IR}^{\rm N6946}$\fi}

\newcommand{\abell}{A1703-zD1}

\newcommand{\la}{\raisebox{-0.5ex}{$\,\stackrel{<}{\scriptstyle\sim}\,$}}
\newcommand{\ga}{\raisebox{-0.5ex}{$\,\stackrel{>}{\scriptstyle\sim}\,$}}

\begin{document}

\title*{Lyman alpha emitting and related star-forming galaxies at high redshift}
\author{Daniel Schaerer}
\institute{Observatoire de Gen\`eve, Universit\'e de Gen\`eve, 51 Ch. des Maillettes, 1290 Versoix, Switzerland;
CNRS, IRAP, 14 Avenue E. Belin, 31400 Toulouse, France;
\email{daniel.schaerer@unige.ch}
}
%
%
\maketitle

\abstract*{I  provide an overview about star-forming galaxies at high redshift and their physical properties.
Starting from  the  populations of  \lya\ emitters and Lyman break galaxies, I summarize their common features and distinction.
Then I summarize recent insight onto their physical properties gained from SED models including nebular emission,
and various implications from these studies on the properties of star-formation at high redshift.
Finally, I present new results and an overview on the dust content and UV attenuation of $z>6$ galaxies obtained 
from IRAM and ALMA observations.}

\abstract{I  provide an overview about star-forming galaxies at high redshift and their physical properties.
Starting from  the  populations of  \lya\ emitters and Lyman break galaxies, I summarize their common features and distinction.
Then I summarize recent insight onto their physical properties gained from SED models including nebular emission,
and various implications from these studies on the properties of star-formation at high redshift.
Finally, I present new results and an overview on the dust content and UV attenuation of $z>6$ galaxies obtained 
from IRAM and ALMA observations.}

\section{Introduction}
\label{s_intro}
The so-called Lyman alpha emitters (LAEs) and Lyman break galaxies (LBGs) make up the dominant population of the 
current known star-forming galaxies at high redshifts, both in terms of numbers and of contributors to the total 
star formation rate (SFR) density of the Universe. 
They constitute one of the main windows to study the early Universe
and address a variety of questions concerning galaxy formation and evolution, star formation at high redshift, the connection
between galaxies and the cosmic web, cosmic reionisation and others.
We will briefly address some of these questions here, providing, however, a non-exhaustive, partial view on LAEs and LBGs, 
and presenting some recent results on dust emission and UV attenuation of high-z star-forming galaxies.
Other types of high redshift galaxies, as found e.g.\ by selection at IR--millimeter wavelengths or GRB host 
galaxies, are not discussed here.

\section{LAEs and LBGs}
\label{s_laelbg}
When discussing the properties, nature, precursors, progenitors etc.\ of LAEs and LBGs it is important to
remember the fundamental selection criteria defining these galaxy populations.
The well-known Lyman break selection, or drop-out technique, targets the UV restframe emission of galaxies using the rapid flux decrease
(``break")  shortward of \lya\ (1216 \AA)  or of the Lyman limit (912 \AA) to chose a specific redshift domain.
To avoid confusion with old stellar populations or very dusty galaxies, a color cut retaining only relatively blue objects
in their rest UV is also adopted, implying
a selection for star-forming galaxies which are not too strongly reddened.
%
The LAE selection  targets galaxies with \lya\ in emission, which are selected through excess in a narrow-band filter with respect
to the nearby continuum. Often this criterium is combined with "drop-out" criteria to
select a specific redshift range and avoid ``contamination" by other emission lines (cf. \cite{ouchi08}). 
As for LBGs, follow-up spectroscopy is frequently carried out, largely confirming the redshift/nature of these galaxies
(cf.\ \cite{rhoads03,Kashikawa06}).

The selection criteria are important, as they will to some extend determine distinctions and overlaps between the 
two galaxy populations. Here, since both LAEs  and LBGs are targeting the same spectral range (i.e.\ rest-frame UV,
generally dominated by emission from young/recent massive star formation) it is evident that they must overlap.
Indeed, numerous findings/results show that this is the case, by how much they overlap, and what distinguishes
them. We shall now summarize the most important arguments.

\begin{figure}[htb]
{\centering
\includegraphics[width=7.5cm]{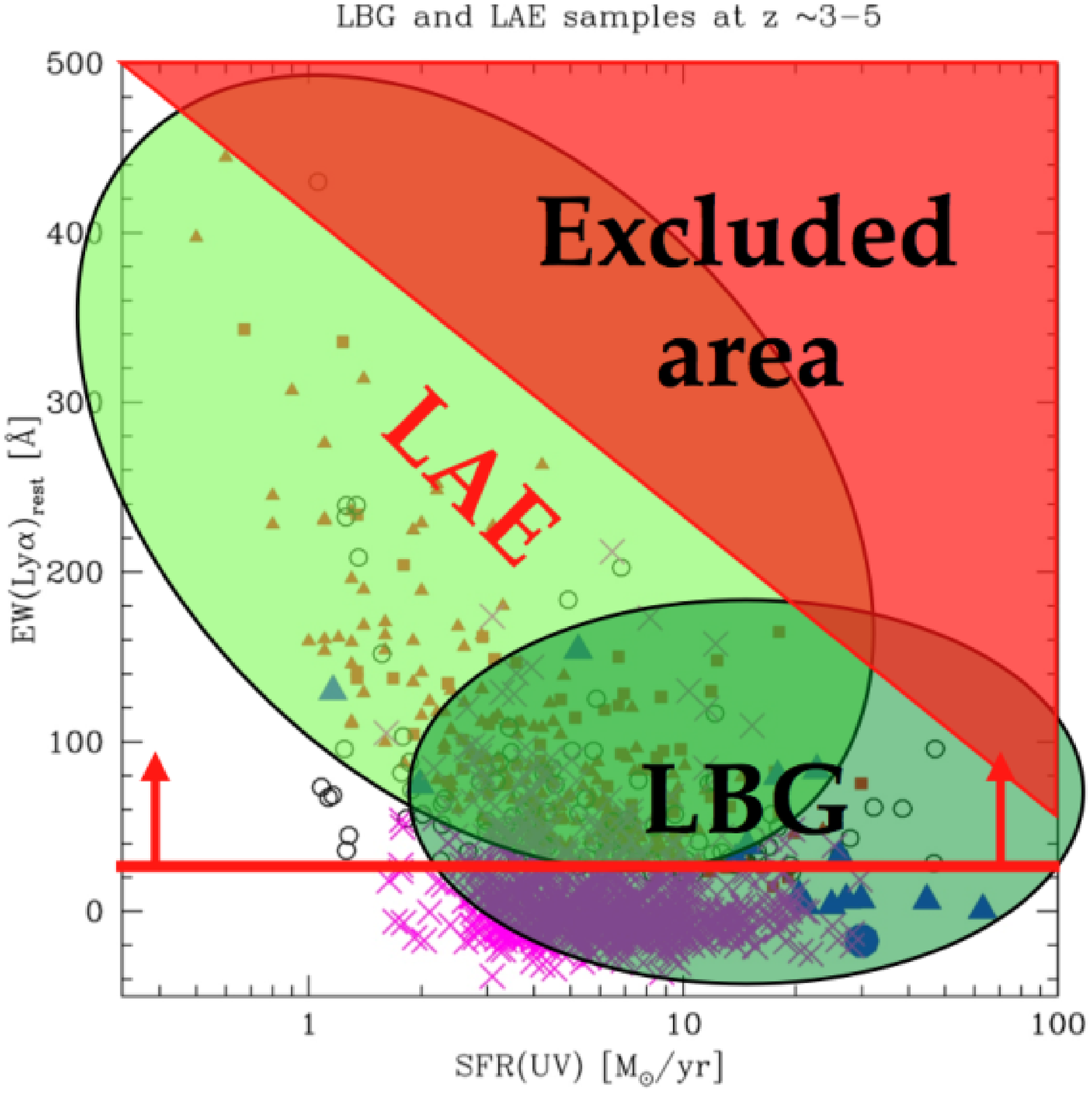}\includegraphics[width=5.5cm]{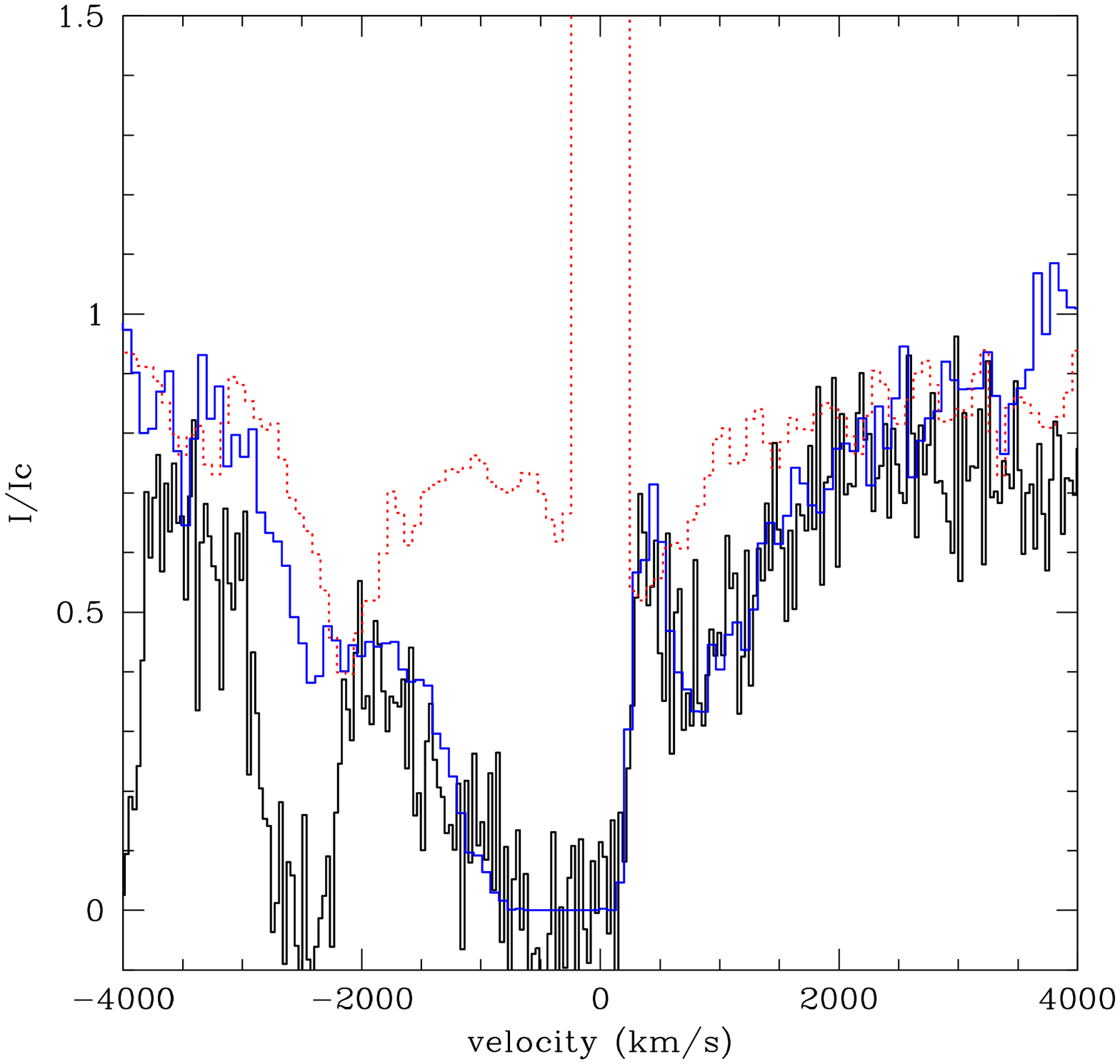}}
\caption{{\bf Left:} \lya\ equivalent width as a function of UV SFR (from the UV magnitude) for LAEs and LBGs at $z\sim 3-4$
showing the overlap between the two galaxy populations.
{\bf Right:} Observed (black solid line) and modeled \lya\ line profile (blue) of the $z=2.7$ LBG cB58 using a radiation transfer
model with outflows and dust. The observed amount of dust attenuation transforms the intrinsic \lya\ emission (dotted line)
into the observed broad absorption line profile. This demonstrates how \lya\ emission from massive star formation
inside this galaxy is changed into absorption after processing by the ISM; i.e.\ how the transition between LAEs and LBGs 
can be understood. From \cite{verhamme08}.}
\label{fig_unify}
\end{figure}

\begin{itemize}
\item A subset of LBGs shows strong enough \lya\ emission (e.g.\ \lya\ equivalent width \wlya $> 20$ \AA) to be selected by 
the traditional narrow-band excess surveys for LAEs. At $z \sim 3$ this corresponds to $\sim 25$ \% of LBGs
\cite{shapley03}, and LAEs restricted to the same continuum magnitude limit are statistically indistinguishable
from these LBGs  (\cite{verhamme08}).
Conversely, LAEs generally also fulfill LBG selection criteria, when data/depth permits.
\item The number density of LAEs increases with redshift, converging towards the same density as LBGs 
at $z \sim 6$ (\cite{ouchi08}). 
Also, the fraction of LBGs showing \lya\ in emission increasing towards higher redshift 
\cite{stark11,schaereretal2011,2012MNRAS.422.1425C},
further direct proof of the increasing overlap between the two populations at high $z$.
\item Observable properties, such as \wlya\ versus UV magnitude, of LAEs and LBGs show the same 
behavior, i.e. the absence of high \wlya\ in UV bright galaxy, a trend often called the ``Ando-plot" (see Fig.\ \ref{fig_unify}).

\item The emergent \lya\ flux (line profile) is strongly dependent on the dust content, and intrinsic \lya\ emission
related to the presence of massive stars (\hii\ regions) in LBGs is readily transformed by radiation transfer
effects into the observed variety of \lya\ profiles including \lya\ absorption \cite{verhamme08,schaerer08}.
This demonstrates how  LAEs and LBGs can be transformed into each other, and suggests that dust is the main physical 
distinction between them (at least on average). Obviously, since the amount of dust attenuation generally
depends on stellar mass, the latter quantity may be a more fundamental, underlying driver.
Indeed lower masses (on average) have been found by many studies for LAEs (e.g.\ \cite{Gawiser07}).
\item Radiation transfer effects also naturally explain the observed changes of the \lya\ properties and LAE/LBG 
populations with redshift, as long as the average dust extinction diminishes towards high-z
\cite{verhamme08,2011ApJ...730....8H,2012MNRAS.422..310G},
which is in line with other observations (\cite{2013A&A...554A..70B}).

\item Correlation lengths of populations of LAEs and LBGs at $z \sim 3$ are comparable and indicate
lower halo masses for LAE 
(cf.\ \cite{Adelberger05,Gawiser07} also  \cite{2013MNRAS.433.2122C}).
 \cite{2013MNRAS.433.2122C} also find an influence of environment (galaxy density) on the 
difference between LBGs with or without \lya\ emission. 

\end{itemize}

In short, the above arguments demonstrate a fundamental overlap between the UV-selected star-forming galaxy populations
of LAEs and LBGs (growing with redshift), and that lower dust content related to lower mass makes the main 
distinction between them (at a given redshift). For most purposes we can therefore consider them as a ``continuum"
of star-forming galaxies in a unified manner.
For other information on these galaxy populations see also the reviews of 
\cite{2011ARA&A..49..525S} and \cite{Dunlop2012Observing-the-f}.

\subsection{Star formation in UV selected galaxies at high redshift}
Numerous studies have recently discussed and revised the physical properties of LBGs and LAEs at high $z$,
such as their stellar mass, SFR, specific SFR, age, dust attenuation, and related quantities.
Indeed, thanks to great progress achieved primarily with HST (the new WFC3 camera in particular), deep ground-based
imaging in complementary bands (e.g.\ U and K), and deep imaging with Spitzer's IRAC camera,
we now have sizable samples of galaxies at $z \sim$ 3--8 with a decent photometric coverage, allowing such studies.
E.g.\ the recent work of \cite{Bouwens2014UV-Luminosity-F} has identified more than 10'000 LBGs at $z>4$
with photometric coverage up to 1.6 $\mu$m at least.

Regarding the derivation of the physical properties of high-z LBGs, a significant advance has been the recognition	
that emission lines (nebular emission) significantly contribute to the broad-band photometry of theses galaxies, and that
their effect must be taken into account 
(\cite{SdB09,schaerer&debarros2010,shimetal2011,2014A&A...563A..81D,Smit2013,Labbe2013,stark2012}).
While the importance of emission lines is now widely established, its precise effect on the physical parameters
remains debated to some extent (cf. e.g.\ \cite{2014A&A...563A..81D,Gonzalez2012,2013ApJ...772..136O}).

Overall implications from the inclusion of nebular emission are manyfold, as discussed in detail
by \cite{schaerer&debarros2010,2014A&A...563A..81D,schaerer2014} and also by \cite{2014A&A...566A..19C}. 
Schematically one finds younger ages, lower stellar masses, higher 
dust attenuation, and a higher specific SFR (rising with increasing redshift), when nebular lines are included in SED fits. 
Furthermore the data seems to favor star-formation histories varying over relatively short timescales, which
are not compatible with the often-made assumption of constant SFR over $>$ 100 Myr.
This latter result implies e.g.\ that standard calibrations, such as the SFR(UV) or SFR(IR) relations
from \cite{kennicutt1998} are not valid, and that the simple relation between color excess and the UV slope 
$\beta$, which is commonly used to determine dust attenuation of high-$z$ galaxies, breaks down
(see discussions in \cite{schaerer2013,2014A&A...566A..19C}).
The preference for variable star-formation histories comes (for $z\sim$ 3.8 to 5 galaxies) from their
broad range of (3.6-4.5) $\mu$m colors, a direct tracer of the \ha\ equivalent width, which is not compatible
with constant SFR and ages $>50$ Myr \cite{2014A&A...563A..81D}. It seems unavoidable to introduce
some variation of SFR (probably some stochastic events), although exponentially declining and rising 
SF histories do a reasonable job and are difficult to distinguish \cite{2014A&A...563A..81D}.
Again, measures of the Balmer break (cf.\ \cite{2013ApJ...772..136O}) and the \ha\ excess (3.6-4.5 color) 
cannot be reconciled with simple models assuming constant SFR (\cite{fournier2014}).
Other support for variable SF histories come from clustering analysis \cite{leeetal2009}, galaxy models including
feedback \cite{Wyithe2013,Hopkins2013}.

\begin{figure}[htb]
\centering
\includegraphics[width=9cm]{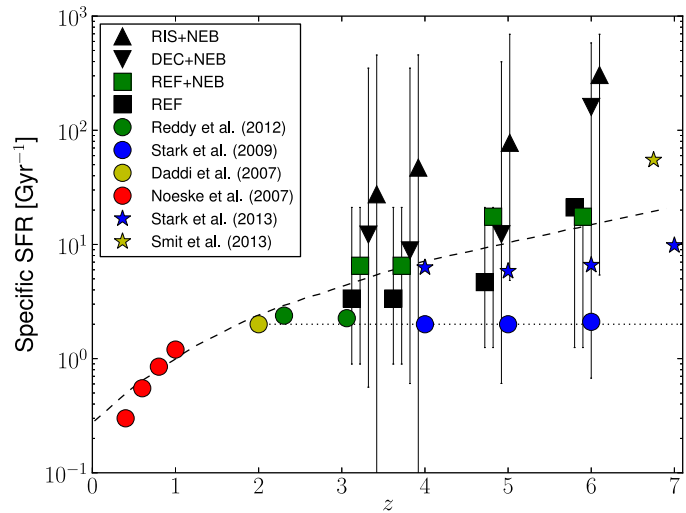}
\caption{Specific star formation rate sSFR as a function of redshift determined for a large sample of LBGs at $z>3$ using
models including  nebular emission. Note the average increase of sSFR(z) and the large scatter. Figure from \cite{2014A&A...563A..81D}.}
\label{fig_ssfr}
\end{figure}

As already pointed out several years ago \cite{schaerer&debarros2010,debarros2011sf2a} SED models
including nebular emission yield quite naturally significantly higher specific star formation rates, sSFR,
than more simple models, and an average sSFR(z) rising with redshift, in better agreement
with predictions from the popular gas accretion-driven galaxy models 
(e.g.\ \cite{boucheetal2010,weinmannetal2011,2013ApJ...772..119L}).
A more recent plot showing sSFR as a function of redshift, determined from a large sample of LBGs
analyzed in a homogeneous fashion including nebular emission and variable SF histories, is shown in 
Fig.\ \ref{fig_ssfr}.

If the star formation histories of distant galaxies vary on relatively short timescales, another 
logical implication is that of (significant) scatter in quantities related to the SFR, i.e.\ in the SFR--mass
diagram, in the sSFR (at a given redshift) etc. See e.g.\ Figure\ \ref{fig_ssfr} and discussions in 
\cite{debarros2011sf2a,schaerer2013,2014A&A...566A..19C}).
Note, however, that different observables, such as the UV or IR luminosity,  or the \ha\ flux, which
all trace the SFR,  respond on different timescales, i.e.\ are expected to show various degrees of scatter
\cite{schaerer2013}. Establishing clearly the amount of scatter, hence variability/stochasticity of star formation
in distant galaxies, and how this fits together with apparently more smooth star formation -- i.e.\ a well-defined
star formation main sequence observed at lower redshifts ($z < 1-2$, e.g.\ \cite{Noeske07,daddietal2007,elbaz2011}) --
remains to be done. Physically, variations of star formation on timescales $<100$ Myr inside high-z
galaxies seem quite natural e.g.\ in view of decreasing timescales (e.g.\ the dynamical timescale) and
strong feedback effects, which must occur in small galaxies with intense star formation. They are
currently not accounted for in the various models describing the evolution of galaxies at equilibrium,
such as the ``bathtub" model.

Among the LBGs from $z \sim 3-6$ \cite{2014A&A...566A..19C} find indications for approximately 2/3 of galaxies showing
more signs of current SF activity (nebular emission), and $1/3$ showing weak or absent emission lines.
This behavior seems present at all UV magnitudes, but both its significance and origin remain uncertain.
Other authors, have suggested  duty cycles for the star-forming galaxies at $z>4$ \cite{Jaacks2012,Wyithe2013},
which, if correct, may imply the existence of currently undetected galaxy populations. 

Clearly, further progress
both observationally and our theoretical understanding of distant galaxies remains to be done.
Even if current accretion-driven galaxy (``bathtub") models provide important insight into
galaxy formation and its connection with the cosmic web, these models represent a simplified picture, which still
lack significant ingredients. From observations, we need more direct and precise measures of their
star formation rate, stellar masses, dust attenuation, star formation histories, ``burstyness" and related quantities,
which will rely both on new observations (e.g.\ with the JWST and with ALMA) and more 
realistic modeling to infer the galaxy properties.

In the next Section, I briefly present recent results from attempts to determine the dust content 
and UV attenuation in high-z galaxies.

\begin{figure}[htb]
{\centering
\includegraphics[width=6cm]{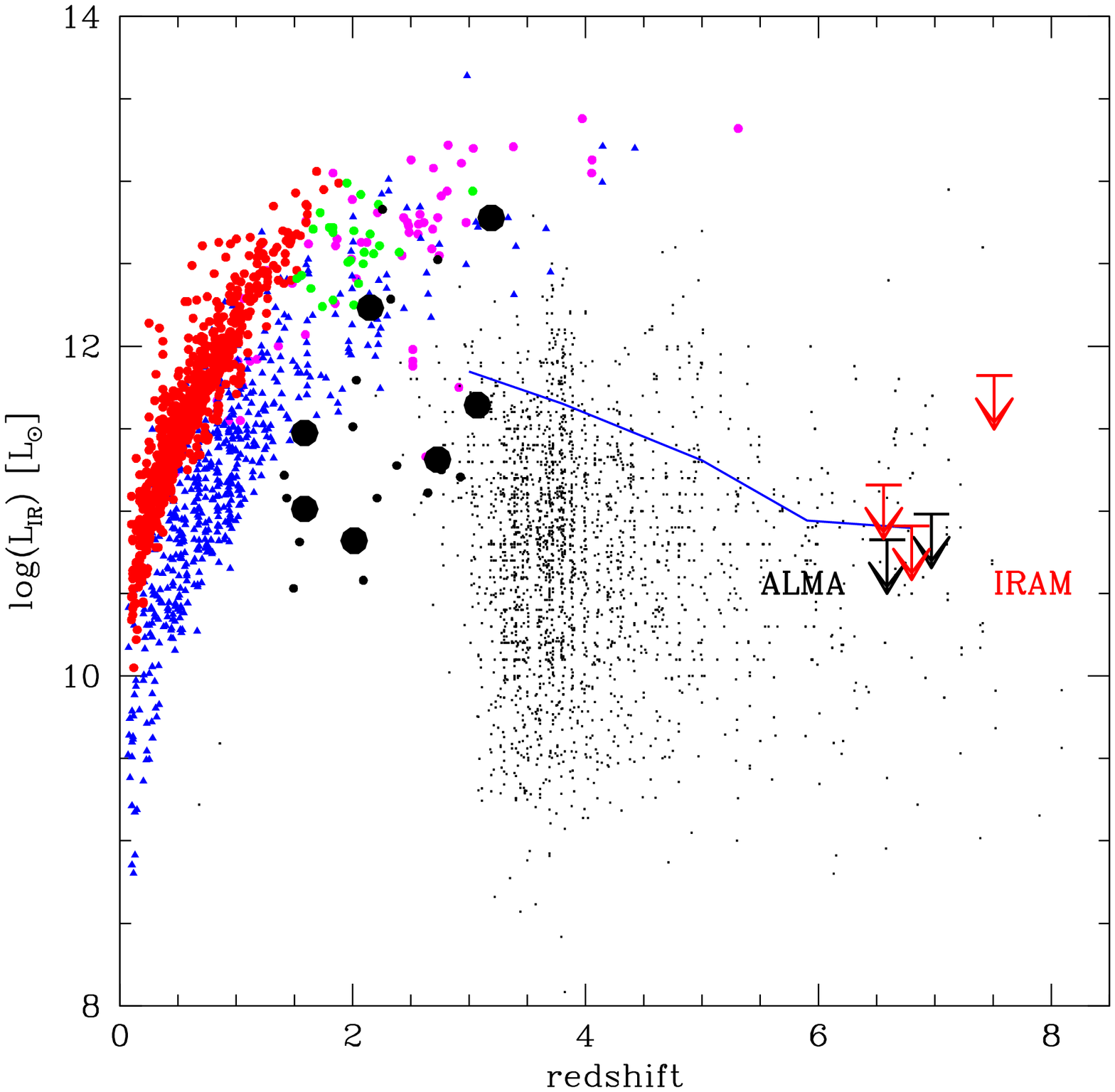}\includegraphics[width=6cm]{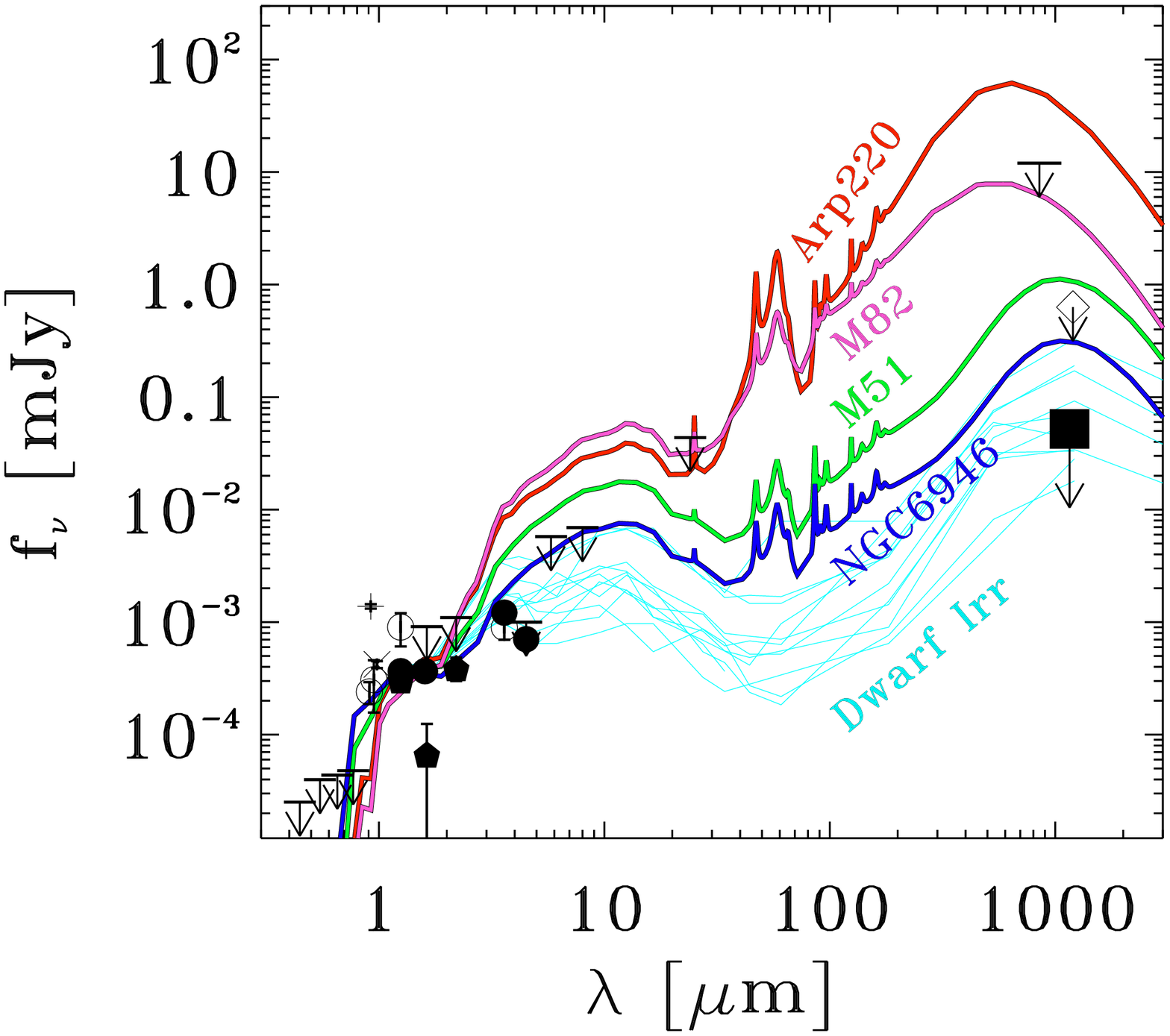}}
\caption{{\bf Left:} IR luminosity (derived for $T_d=35$ K, taking the CMB into account) versus redshift for the objects 
discussed in this paper (arrows at $z>6.5$) and for other samples.
Colored  small circles show galaxies detected with  HERSCHEL in various blank fields
(red, \cite{2013MNRAS.431.2317S}; blue \cite{2011A&A...533A.119E}, green: \cite{2010MNRAS.409...22M}).
Large and small black circles show the lensed galaxies studies by \cite{2014A&A...561A.149S}
and \cite{2013ApJ...778....2S} respectively.
The small dots show the predicted IR luminosity of LBGs from the sample studied by \cite{2014A&A...563A..81D}  and
\cite{schaerer2014}; the blue line shows the median \lir.
{\bf Right:} Observed SED of HIMIKO covering the optical, near-IR, to IR/mm domain and comparisons to the SEDs of local/nearby
galaxies. Taken from \cite{2013ApJ...778..102O}.
}
\label{fig_lir_z}
\end{figure}

\section{Dust and UV attenuation in $z>6$ galaxies and comparisons to low redshifts}
\label{sec:2}
Using the GISMO 2mm camera and Widex 1.2mm observations with the Plateau de Bure interferometer 
we have recently targeted to $z\ga7$ galaxies, the strongly lensed LBGs \abell\ with a well-defined photometric
redshift $z \approx 7.0$ discovered by \cite{2012ApJ...747....3B}.  and the spectroscopically confirmed $z=7.508$ LBG 
from \cite{2013Natur.502..524F}. Both galaxies were non-detected with an rms of 0.12-0.17 mJy/beam in the 
continuum at 1.2mm, and their \Cii\ emission was not also not detected. 
This new data, together with recent ALMA and IRAM observations of three other $z>6$ galaxies
(the LAEs named Himiko, IOK-1, and HCM6A \cite{}), are analyzed in a homogenous way in \cite{schaerer2014_iram},
from which we here show some preliminary results.

The upper limits on the IR luminosity determined from this data are shown in Fig.\ \ref{fig_lir_z}, where
we also plot \lir\ measurements with Herschel at lower redshift ($z$ \la\ 3), and the predicted IR luminosities
of LBGs from the sample of \cite{2014A&A...563A..81D}. As expected, the IR luminosity limits of strongly lensed galaxies
observed with IRAM (\abell, HCM6A) reach comparable effective depth as recent ALMA data
for blank field (unlensed) galaxies. They are reaching luminosities $\lir$ \la\ $10^{11}$ below the LIRG regime.

The SED of one of these galaxies, here HIMIKO, is show in the right of Fig.\ \ref{fig_lir_z}.
Clearly, the SED of this object shows comparatively significantly less IR emission than that of ULIRGs, such as Arp220 and M82, 
and than spiral galaxies. It is more compatible with that of nearby dwarf galaxies, a results which also
holds for the other four $z>6$ galaxies included in our sample, as discussed previously by 
\cite{2007A&A...475..513B,2012ApJ...752...93W,2013ApJ...771L..20K,2013ApJ...778..102O,Ota2014ALMA-Observatio}).

\begin{figure}[htb]
\centering
\includegraphics[width=7cm]{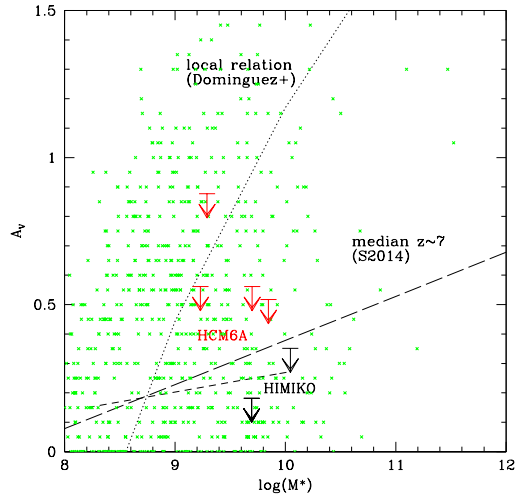}
\caption{Visual attenuation, $A_V \approx \auv /2.5$, for the objects of this study as a function of stellar mass.
Small green dots show the values determined by \cite{2014A&A...563A..81D}  for a large sample of $z\sim 4$ LBGs.
The dotted line shows the mean relation for local star-forming galaxies from \cite{dominguez2013}.
Two median relations for LBGs at $z\sim 7$ from \cite{schaerer2014} are shown 
as long- and short-dashed lines respectively.}
\label{fig_mstar_av}
\end{figure}

\begin{figure}[htb]
\centering
\includegraphics[width=7cm]{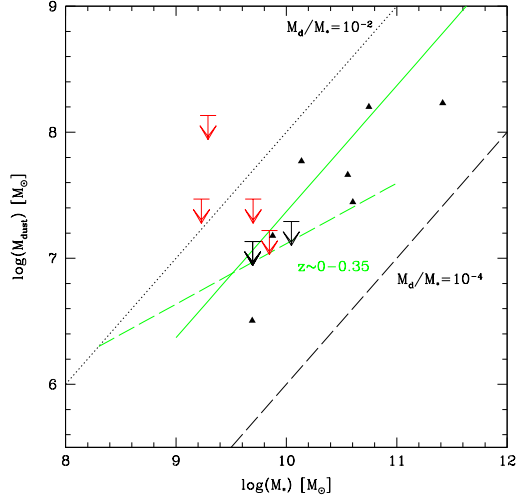}
\caption{Dust mass as a function of stellar mass for our objects (upper limits) and other related galaxies.
The typical uncertainty on \mdust\ due to the unknown dust temperature is $\sim \pm 0.4$ dex.
Strongly lensed galaxies at $z \sim$ 1.5--3 from \cite{2013ApJ...778....2S,2014A&A...561A.149S}, probing 
a similar mass range, are shown as black triangles and squares respectively. 
Solid, dashed  lines: $\mdust/\mstar=10^{-2}$, $10^{-4}$  (\mstar\ here assumes Salpeter IMF).
The green dashed line shows the location of the sequence observed by the H-ATLAS/GAMA survey at
	$z \sim$ 0--0.35 \cite{Bourne2012Herschel-ATLAS/}; the green solid line the median value of $M_d/\mstar=-2.63$ obtained
	by \cite{Smith2012} from the H-ATLAS survey after adjustment to the Salpeter IMF used here.
	The dust-to-stellar mass ratio of the high-z galaxies studied here is compatible with values observed at
	lower redshift, down to the nearby Universe. 
		}
\label{fig_mdust}
\end{figure}

From the ratio of the IR/UV luminosity it is straightforward to determine the upper limits on dust attenuation
in these galaxies.
From the nearby Universe out to $z \sim 2$ and possibly higher, various measurements
(Balmer decrement, IR/UV, and others) yield a correlation between the dust attenuation
and stellar mass, which apparently also shows little or no evolution with redshift 
\cite{2013ApJ...763..145D,2012ApJ...754L..29W,2010A&A...515A..73S}.
Even LBG samples at $z \sim 3-7$ show such a correlation 
\cite{2010A&A...515A..73S,2014A&A...563A..81D,schaerer2014}
although these may obviously affected by selection effects and biases.
It is therefore interesting to examine the constraints placed by the new UV attenuation
data as a function of the galaxy mass. This is shown in Fig.\ \ref{fig_mstar_av},
where we also plot the mean relation derived at low redshift,
the values derived from SED fits for $z \sim 4$ LBGs, and the median relation
for $z \sim 6.8$ LBGs using the same SED fitting procedure.

The upper limits for several of our galaxies fall  below the 
local relation, indicating less dust attenuation than would be expected
on average for $z \sim 0$ galaxies with the same stellar mass. On the other hand,
the limits on dust attenuation are in good agreement or do not deviate strongly
from the median relation found from our modeling of a sample of 70 LBGs with a median
$z_{\rm phot}=6.7$ (SdB14). Although the upper limit for Himiko deviates most from
this relation, we do not consider this as discrepant with expectations. Indeed, 
from SED modeling one also finds relatively massive galaxies  with low attenuation
(cf.\ \cite{2014A&A...563A..81D}),
as probably also corroborated by the empirical finding of an increasing 
scatter of the UV slope towards brighter magnitudes (cf.\ \cite{Bouwens09_beta}).

Finally, the non-detections at 1.2mm provide information on the dust mass of the $z>6$ galaxies.
In Fig.\ \ref{fig_mdust} we plot the limits on dust mass as a function of stellar mass and compare this
to average values found in the nearby Universe and at higher redshift. The present limits on dust mass are 
not  incompatible with the standard dust-to-stellar mass ratios observed at low redshifts. They also agree with the
recent observations of strongly lensed galaxies at $z\sim 1-3$ detected with Herschel.
In short, we conclude that the available data for high redshift star-forming galaxies shows no significant evolution
from $z \sim 0$ to 3 and remains compatible  with this out to the highest redshifts currently probed.
The current data does not show evidence for a downturn of \mdust/\mstar\ at high redshift, in contrast e.g.\ to
 the claim by \cite{Q.Tan2014Dust-and-gas-in}.
As already mentioned before, deeper observations and larger samples are needed to determine the 
evolution of dust with redshift, galaxy mass, and other important parameters.

\begin{acknowledgement}
I wish to warmly thank the organizers, David Block, Ken Freeman, and Bruce Elmegreen for this very interesting and 
exceptional meeting. I thank my collaborators, in particular Anne Verhamme, Frederic Boone, 
Mirka Dessauges, Michel Zamojski, Stephane de Barros, and Panos Sklias for their contributions to some of the work
presented here.
\end{acknowledgement}
%

\bibliographystyle{spphys}
\bibliography{merge_misc_highz_literature}

\end{document}